\documentclass[11pt]{article}
\usepackage{moriond,epsfig}

\def\be{\begin{equation}}
\def\ee{\end{equation}}
\def\bea{\begin{eqnarray}}
\def\eea{\end{eqnarray}}

\begin{document}
\vspace*{4cm}
\title{Searches for new physics at HERA}

\author{Johannes Haller}

\address{Physikalisches Institut, Universit\"at Heidelberg\\
Philosophenweg 12, 69120 Heidelberg, Germany\\
(on behalf of the H1 and ZEUS collaborations)}

\maketitle\abstracts{
The H1 and ZEUS collaborations have searched for signals of physics beyond the Standard Model in $e^{\pm}p$ collisions at a center-of-mass energy of $301-319$\,GeV. 
During the HERA\,I phase each experiment accumulated an integrated luminosity of about 110\,$\mathrm{pb}^{-1}$ for $e^+p$ and about 15\,$\mathrm{pb}^{-1}$ for $e^-p$ collisions.    
The data have been analysed searching for contact interaction effects, leptoquarks, lepton flavor violation, squarks in $R$-parity violating supersymmetric models, excited fermions and FCNC single top production. In this article a summary of the results is given.
}

\section{Introduction}
At HERA electrons or positrons of $27.6$\,GeV and protons of $920$\,GeV ($820$\,GeV before 1998) are collided, resulting in a center-of-mass energy $\sqrt{s}$ of $319$\,GeV ($301$\,GeV).
In these collisions $eq$ interactions are probed at high energies.
The integrated luminosity collected by each experiment during the HERA\,I phase reached roughly $110\,\mathrm{pb}^{-1}$ in $e^+p$ collisions and $15\,\mathrm{pb}^{-1}$ in $e^-p$ collisions.
The maximum square momentum transfer ($Q^2$) reaches $3\cdot 10^4\,\mathrm{GeV}^2$, giving a spatial resolution of $10^{-16}$\,cm.
HERA is particularly sensitive to new particles which couple to an electron and a quark, e.g. leptoquarks or squarks in $R_p$ violating SUSY.
New physics at a scale $\Lambda\gg\sqrt{s}$ may be observable as deviations from the SM in the region of high $Q^2$. 

\section{DIS at high $\mathbf {Q^2}$}
At HERA the dominant process for deep-inelastic $ep$ scattering (DIS) is given by the $t$-channel exchange of a photon between the incoming lepton and a quark from the proton. At low momentum transfer the exchange of massive gauge bosons ($Z^0,W^{\pm}$) is suppressed by propagator terms. 
However at high $Q^2$ the $Z^0$ and $W^{\pm}$ contributions become important. This allows the investigation of electroweak effects in $eq$ interactions.

\begin{figure}[t]
\begin{center}
\epsfig{file=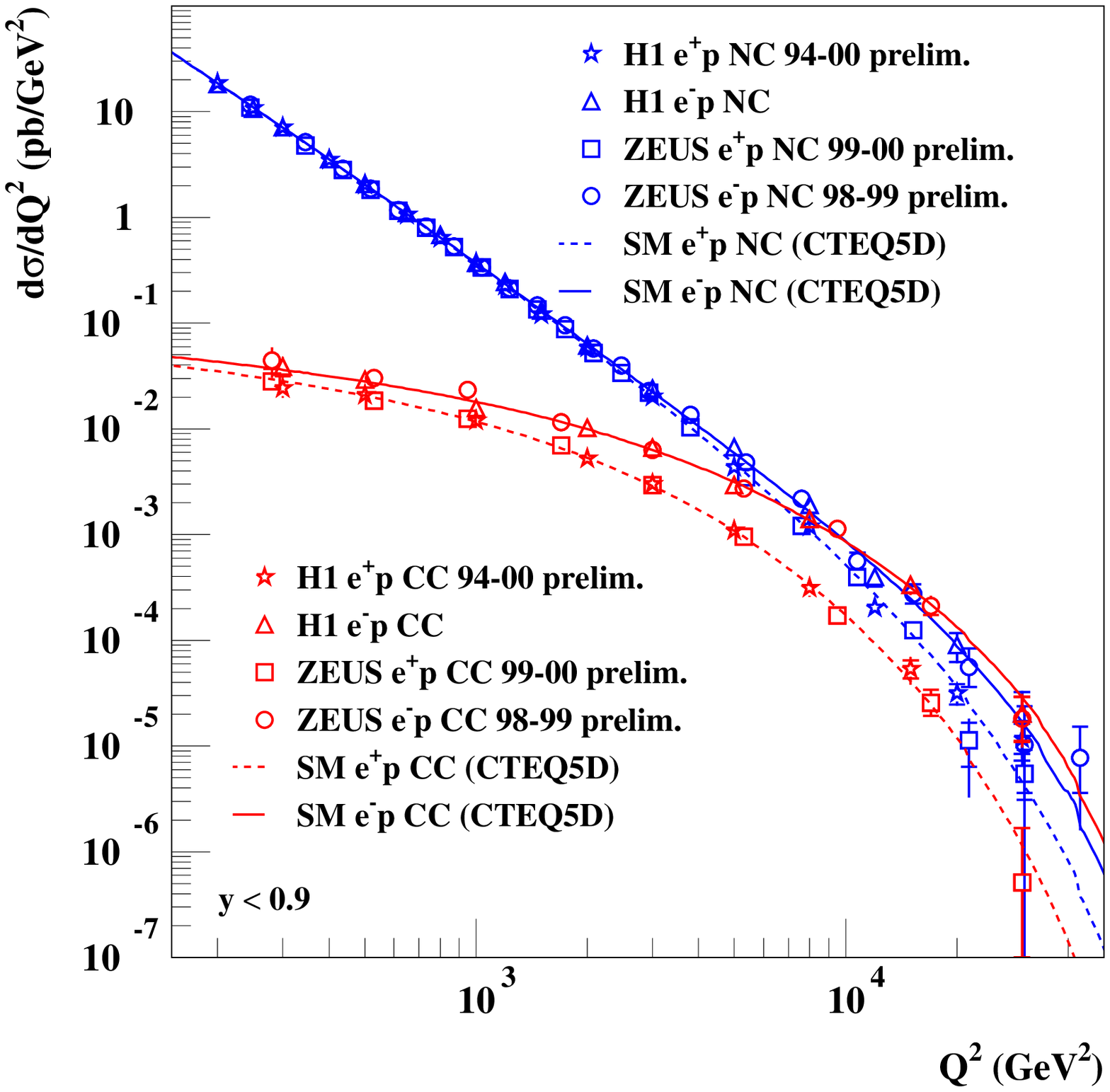,width=7.9cm}
\epsfig{figure=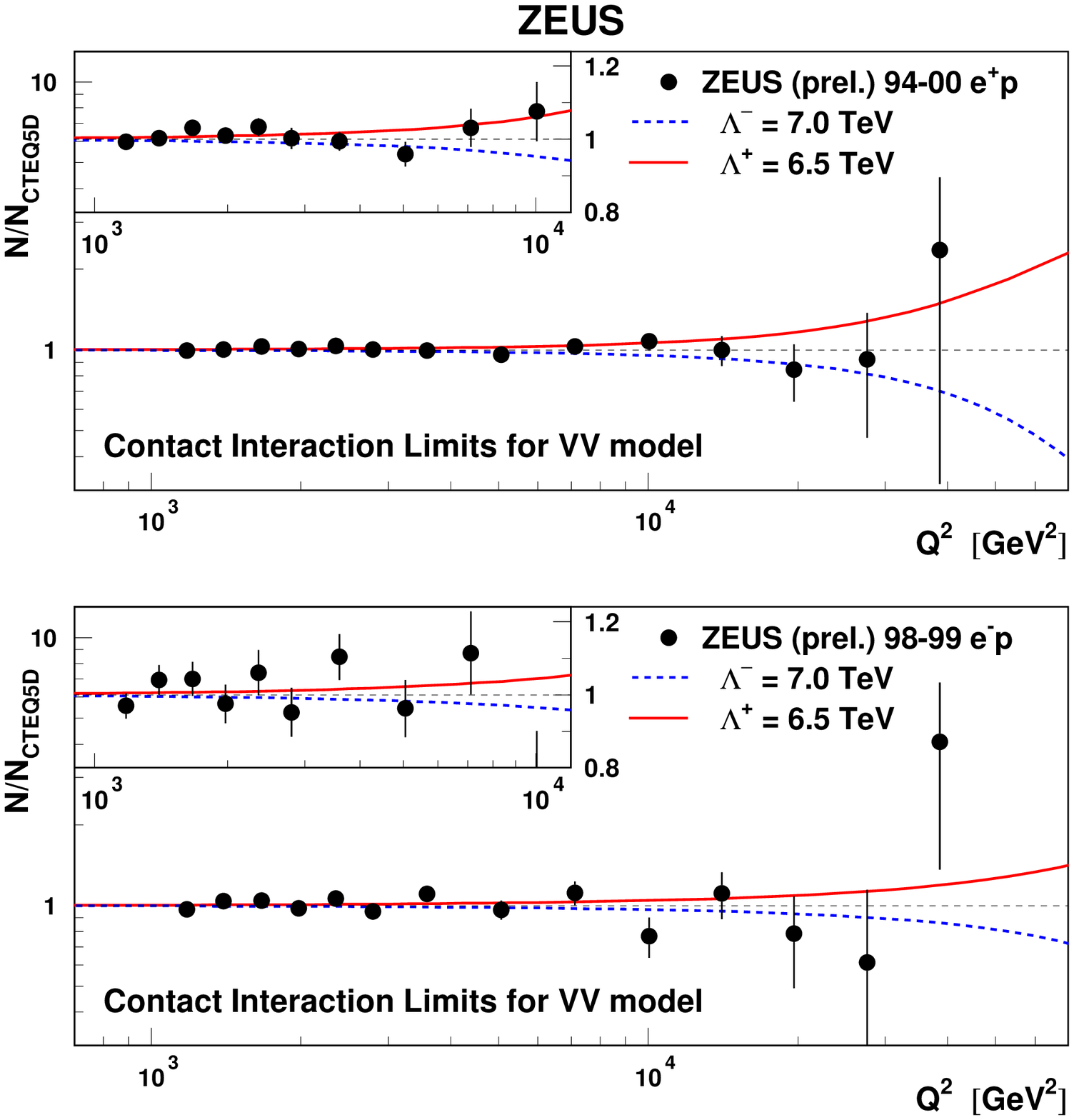,height=7.7cm,clip=}
\caption{(left) Measurements of the differential cross-sections for neutral current (blue) and charged current (red) reactions in deep inelastic scattering at HERA. (right) Ratio of observed number of events over the expectation in neutral current DIS as a function of $Q^2$ as measured in $e^+p$ and $e^-p$ scattering at ZEUS. The fitted curves show the $95$\,\% CL exclusion limits on contact interactions (VV-type) for constructive and destructive interference with the SM process. }\label{fig:nccc}
\end{center}
\end{figure}

In figure~\ref{fig:nccc}\,(left) the H1 and ZEUS measurements~\cite{h1nccc,znccc} of the differential neutral current (NC) and charged current (CC) cross sections $d\sigma/dQ^2$ are shown together with the Standard Model (SM) expectation.
 With the statistics collected at HERA up to the year 2000, the measurements clearly reveal the dependence on the lepton beam charge that is predicted by the SM, i.e. an increased NC cross section for $e^-p$ with respect to $e^+p$ scattering by virtue of positive ($e^-p$) or negative ($e^+p$) interference between $\gamma$ and $Z^0$ exchange.

\section{Contact Interactions}
The virtual exchange of a new heavy boson could produce deviations of the observed NC DIS cross section at high $Q^2$ due to interference with the $\gamma$ and $Z^0$ fields of the SM.
For particle masses and scales well above the available energy $\Lambda\gg\sqrt{s}$, such indirect signatures can be investigated by searching for four-fermion point-like contact interactions (CI).
The most general chiral invariant Lagrangian for neutral current vector-like contact interactions can be written in the form~\cite{cilang} ${\cal L}_V=\sum_{a,b=L,R}\eta_{ab}^q(\overline{e}_a\gamma^{\mu}e_a)(\overline{q}_b\gamma_{\mu}q_b)$. $\eta_{ab}^q=\epsilon\frac{g^2}{(\Lambda^q_{ab})^2}$ are model dependent coefficients of the new process, $g$ is the coupling constant, $\Lambda_{ab}^q$ is the effective mass scale and $\epsilon=\pm1$ is a parameter determining the sign of the interference with the SM.   

The high $Q^2$ neutral current DIS measurements performed by both collaborations~\cite{h1nccc,znccc} show no deviations from the SM expectation. Hence limits on the CI scale $\Lambda$ have been derived for the most common CI models~\cite{h1ci,zci}. 
In figure~\ref{fig:nccc}\,(right) the cross section ratio for data and SM expectation of measured by ZEUS is shown for the $e^+p$ (top) and $e^-p$ (bottom) data together with fits to the vector-vector-type (VV) CI scenario ($\eta_{LL}=\eta_{LR}=\eta_{RL}=\eta_{RR}$) corresponding to $95$\,\% CL exclusion limits. The limits for the various models range up to $7$\,TeV and are comparable to those obtained at LEP and TeVatron studying reactions complementary to HERA.

Similarly, limits can be obtained for models with Large Extra Dimensions~\cite{arkani} (LED). 
In such a model particles can propagate in the ordinary 4-dimensional space, whereas gravitons live in a world with $n\ge2$ extra compactified dimensions.
This model solves the hierarchy problem of the SM by introducing an effective Planck scale $M_s$. This scale is related to the ordinary Planck scale $M_p$ by $M_p^2=R^nM_s^{2+n}$, where $R$ is the size of the extra dimensions. Gravitons are expected to appear in the $(3+1)$-dimensional world as towers of Kaluza-Klein modes and gravitation is a strong force at short distances.
Since gravitation has been tested directly only to the millimetre scale, it is attractive to test LED models at HERA or other collider facilities.
In the LED Lagrangian an additional term $\propto\epsilon/M_s^4$ arises, which accounts for the graviton exchange. 
From a fit to the NC data $95$\,\% CL exclusion limits on the effective scale  $M_s$  have been set by H1 and ZEUS~\cite{h1ci,zci}. They are of the order of $0.8$\,TeV.

\section{Leptoquarks}
\begin{figure}[t]
\begin{center}
\epsfig{figure=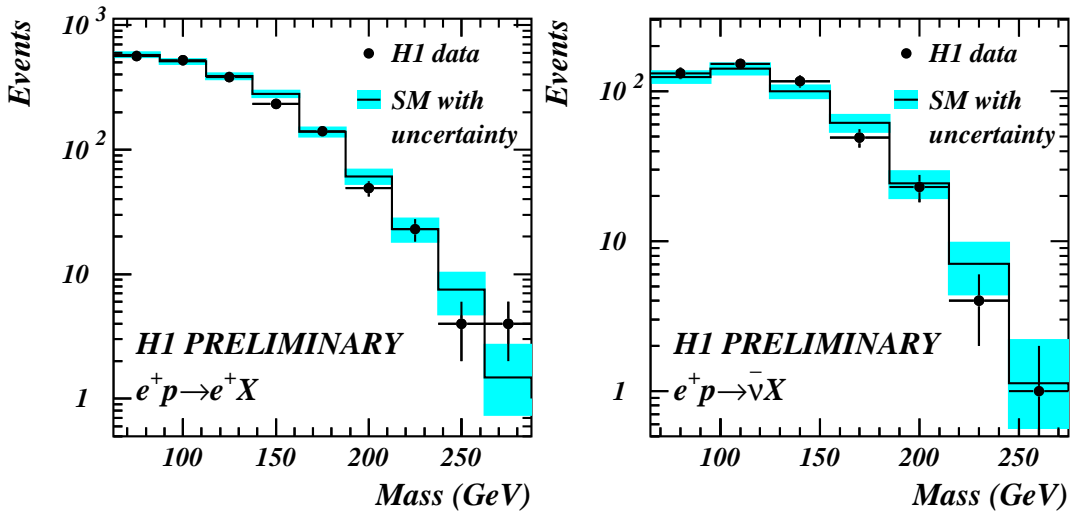,height=7.cm,clip=}
\caption{Invariant mass distribution for $eq$ and $\nu q$ final states for the full $e^+p$ data set.}\label{fig:h1lq}
\end{center}
\end{figure}
Leptoquarks (LQ) are color triplet bosons of spin $0$ or $1$, carrying lepton ($L$) and baryon ($B$) number and fractional electric charge. They couple to lepton-quark pairs and appear in many extensions of the SM which unify leptons and quarks in the framework of Grand Unified Theories (GUT). 

Leptoquarks with masses $M_{LQ}<\sqrt{s}$ could be directly produced at HERA by fusion of the initial state electron and a quark from the incoming proton. The cross section depends on the unknown Yukawa coupling $\lambda$ at the electron-quark-leptoquark vertex.
The $u$-channel exchange of a leptoquark and its interference with the corresponding SM process gives access to masses above the center-of-mass energy of HERA.

The commonly used model of Buchm\"uller, R\"uckl and Wyler~\cite{brw} classifies 14 possible scalar or vector leptoquarks with fermionic number $F=-(3B+L)=0$, $2$.
In this model the branching ratios $\beta_e$ and $\beta_{\nu}$ for the leptoquark decays to $eq$ and $\nu q$ have fixed values ($1,\frac{1}{2},0$).
Leptoquarks with $F=0$, which couple simultaneously to one fermion and one antifermion, are better probed in $e^+p$ collisions due to the large valence quark densities of the proton. 
In contrast, the seven $F=2$ leptoquarks, which couple to two fermions or two antifermions, are better probed in $e^-p$ collisions.

The final states of leptoquark decays and SM DIS reactions are the same.
But the polar angular distribution is different.
Scalar leptoquarks produced in the $s$-channel decay isotropically in their rest frame leading to a flat $d\sigma/dy$ distribution, where $y=\frac{1}{2}(1+\cos\theta^*)$ is the Bjorken scattering variable ({\it inelasticity}) and $\theta^*$ is the polar angle of the lepton in the LQ center-of-mass frame. 
In contrast, for vector leptoquark production the events  would be distributed according to  $d\sigma/dy\propto(1-y)^2$. Both $d\sigma/dy$ spectra are markedly different from the $d\sigma/dy\propto y^{-2}$ distribution expected for the dominant $t$-channel photon exchange in neutral current DIS events. These differences are used to enhance the LQ signal over the DIS background.

The $s$-channel production of LQ would lead to a resonance peak in the distribution of the invariant masses of the decay products.
The invariant mass spectra for $eq$ and $\nu q$ final states of the H1 analysis~\cite{h1lq} are shown in figure~\ref{fig:h1lq} for the full $e^+p$ data set.
No evidence of a signal has been found by either collaboration~\cite{h1lq,zlq}. 
The excess observed in neutral current DIS $e^+p$ data of 1994-1997 at invariant masses of about $200$\,GeV is not confirmed by the new data.

\begin{figure}[t]
\begin{center}
\epsfig{figure=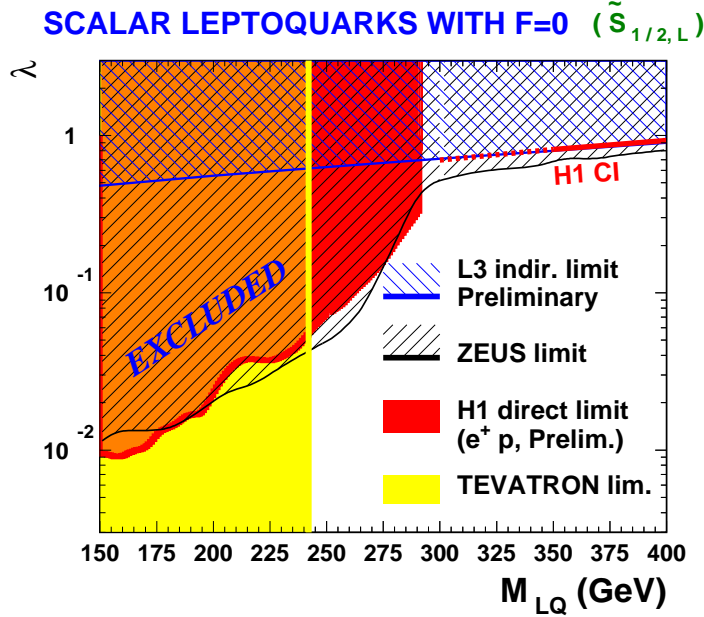,bbllx=10,bblly=120,bburx=540,bbury=600,height=8.5cm}
\caption{Limits on the Yukawa coupling $\lambda$ as a function of the leptoquark mass $M_{LQ}$.}\label{fig:lqlim}
\end{center}
\end{figure}

Both collaborations have derived limits on the Yukawa coupling $\lambda$ as a function of the LQ mass for all LQ types. As an example, the results for a scalar leptoquark with $F=0$ ($\tilde{S}_{1/2,L}$) is shown in figure~\ref{fig:lqlim}. 
At HERA direct limits can be obtained up to the center-of-mass energy. For higher masses the $u$-channel gives sensitivity.
For comparison, the indirect limit obtained from the hadronic cross section in $e^+e^-$ collisions at LEP and the results from the search for direct leptoquark pair production at TeVatron are shown.

In a more general LQ model the branching ratios $\beta_e$ and $\beta_{\nu}$ are treated as free parameters. In this case limits obtained by combining the NC and CC channel are almost independent of the branching ratios~\cite{h1lq} and the comparison with TeVatron limits shows that the HERA results are particularly stringent for small $\beta_e$.

\section{Squarks in $\mathbf{R}$-parity violating supersymmetry}
\begin{figure}[t]
\begin{center}
\epsfig{file=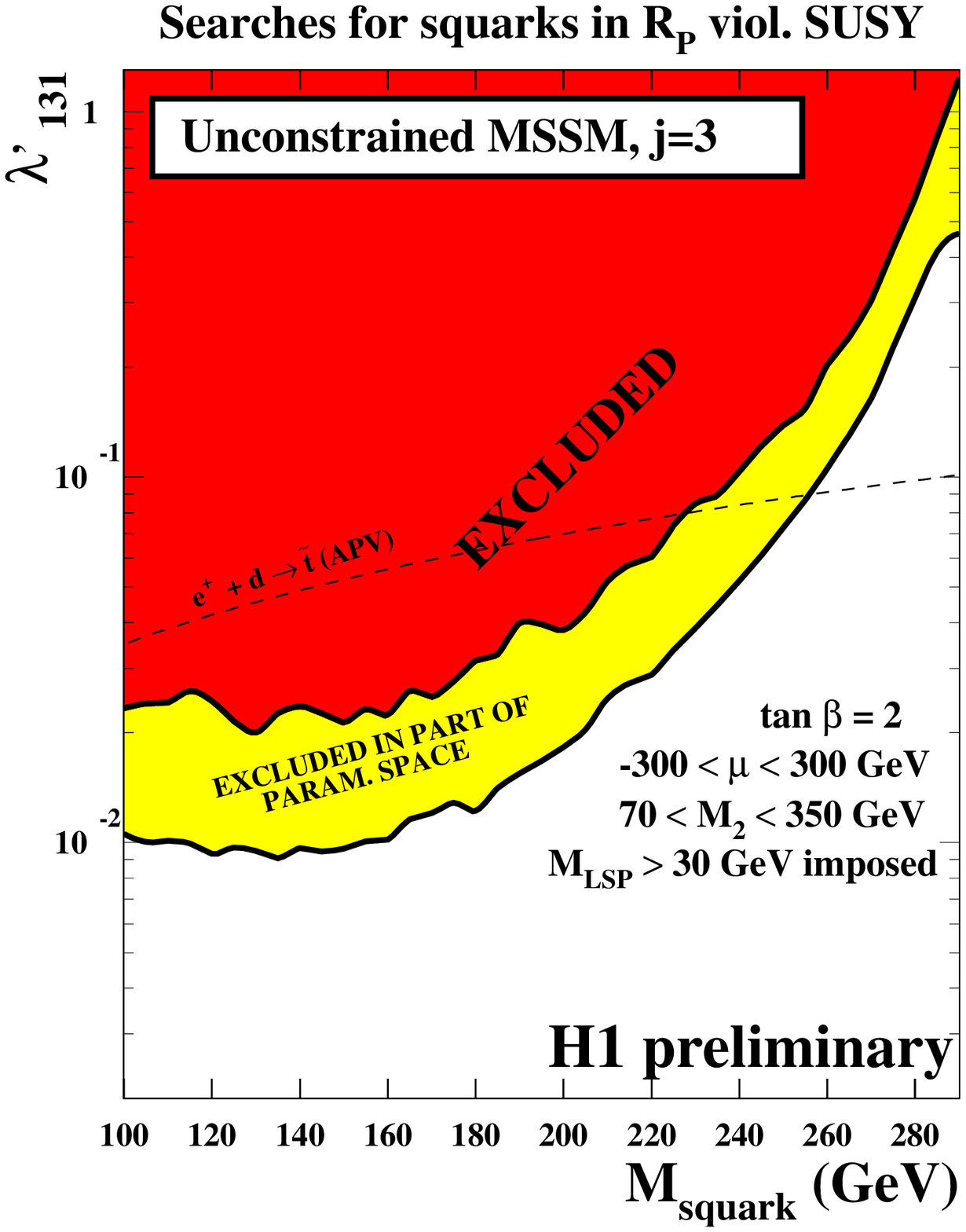,width=7.0cm}
\epsfig{file=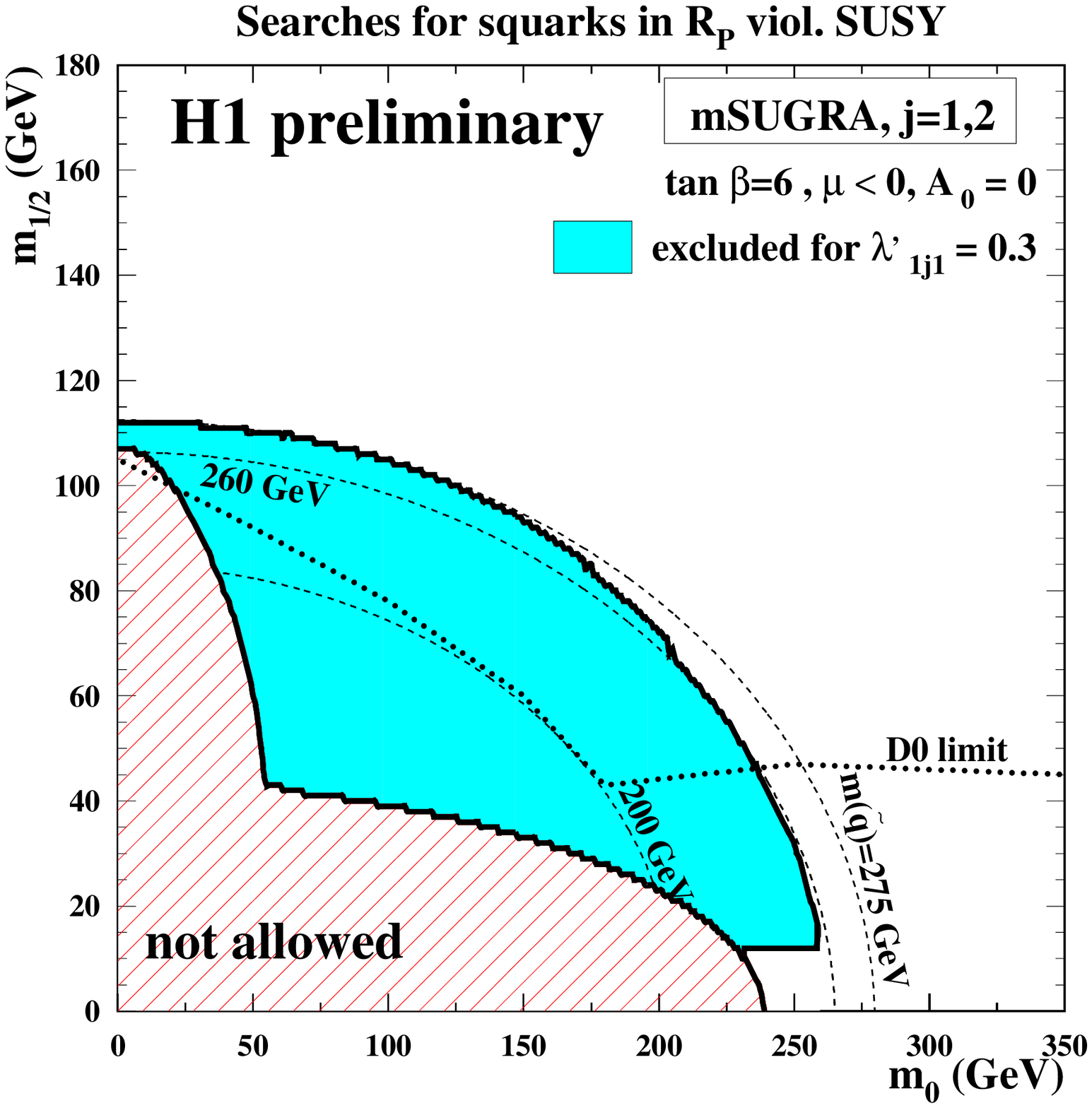,width=7.7cm}
\caption{(left) Exclusion limits on the Yukawa coupling $\lambda'_{131}$ as a function of the squark mass. In order to test the model dependence a scan has been performed in the SUSY parameter space. (right) Exclusion area in the mSUGRA framework for a fixed value of $\tan\beta$. The HERA limits follow squark mass iso-curves and they are $\lambda'$ dependent.}\label{fig:susy}
\end{center}
\end{figure}
The most general SUSY theory which preserves the gauge invariance of the Standard Model allows for Yukawa couplings between two known SM fermions and a squark or a slepton.
Such couplings induce the violation of $R$-parity, which is an multiplicative quantum number. 
Its definition is given by $R_p=(-1)^{L+3B+2S}$ where $L$, $B$ and $S$ are leptonic number, baryonic number and  the spin of a particle.  
Hence $R_p$ is equal to $1$ for particles and equal to $-1$ for sparticles. 
Of special interest at HERA are Yukawa couplings between a squark and a lepton-quark pair~\cite{dreiner} which are described in the superpotential by $\lambda'_{ijk}L_iQ_jD_k$, with $L$ and $Q$ being the left-handed lepton and quark doublet superfields, $i,j,k$ being generation indices and $\lambda'$ being the Yukawa coupling.
A non-zero value for $\lambda'$ would imply the possibility to produce a single squark via $eq$ fusion in the $s$-channel at HERA. 

In models with $R_p$ violation the lightest supersymmetric particle decays to ordinary SM particles.
The squark can decay via two types of processes. The first is the $R_p$ violating decay into an electron or neutrino and a quark, which is called {\it leptoquark-like}. 
The second mode is a so called {\it gauge decay} where the squark decays first into a quark and a gaugino ($\chi_i^0,\chi_i^{\pm},\tilde{g}$). The latter decays $R_p$ violatingly to a lepton and two quarks or to a lighter gaugino and a SM gauge boson. 
Consequently the search for squark production in $R_p$ violating models at HERA has to consider a large variety of final states~\cite{h1susy}. 
Some of these cascade decays lead to very striking signatures with multiple leptons or a lepton with another charge than the initial lepton beam. Here the SM background is negligible.
The sum of the branching ratios of decay channels considered in this analysis is close to one.

All channels have been checked and no deviation from the SM expectation has been observed and limits on the Yukawa coupling $\lambda'$ as a function of the squark mass have been derived. 
Figure~\ref{fig:susy} (left) shows the results interpreted in the {\it phenomenological} minimal supersymmetric extension of the SM (MSSM) where the gaugino masses depend on the MSSM soft terms, but the sfermion masses are free.
A scan in the SUSY parameter space has been performed to check the model dependence of the result.
For a Yukawa coupling of electromagnetic strength ($\lambda'=0.3$) squark masses up to $275$\,GeV are excluded.

In addition the result is interpreted in the framework of minimal Supergravity (mSUGRA)~\cite{msugra}, where a common sfermion (gaugino) mass $m_0$ ($m_{1/2}$) at the GUT scale is assumed.
The electroweak symmetry breaking is driven by radiative corrections.
The excluded area in the ($m_{1/2},m_0$)--plane is shown in figure~\ref{fig:susy} (right) for a fixed value of $\tan\beta$.
The $\lambda'$ dependent H1 limit follows roughly the squark mass isocurve.

\section{Excited fermions}
\begin{figure}[t]
\begin{center}
\epsfig{figure=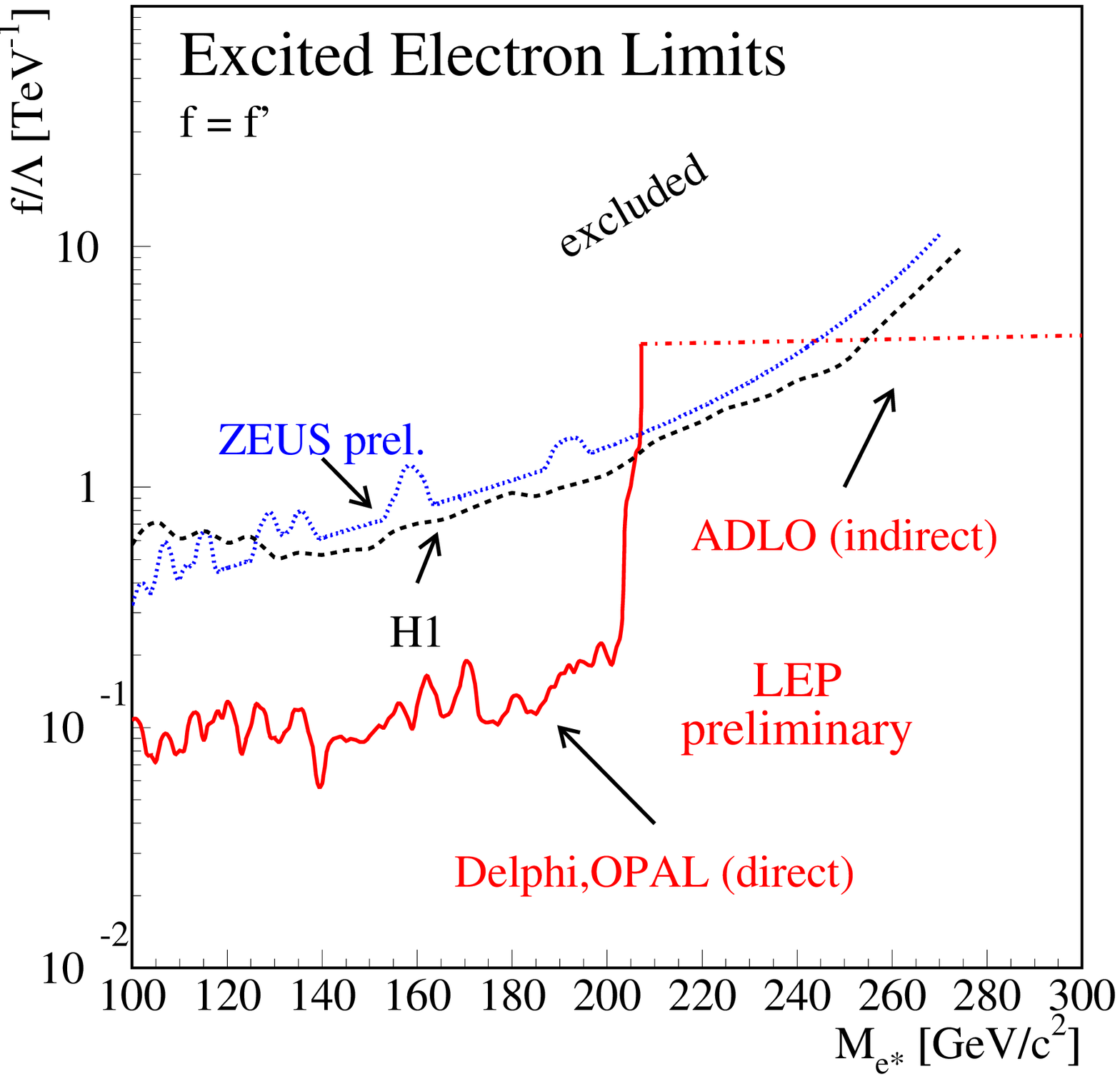,height=7.8cm}
\epsfig{figure=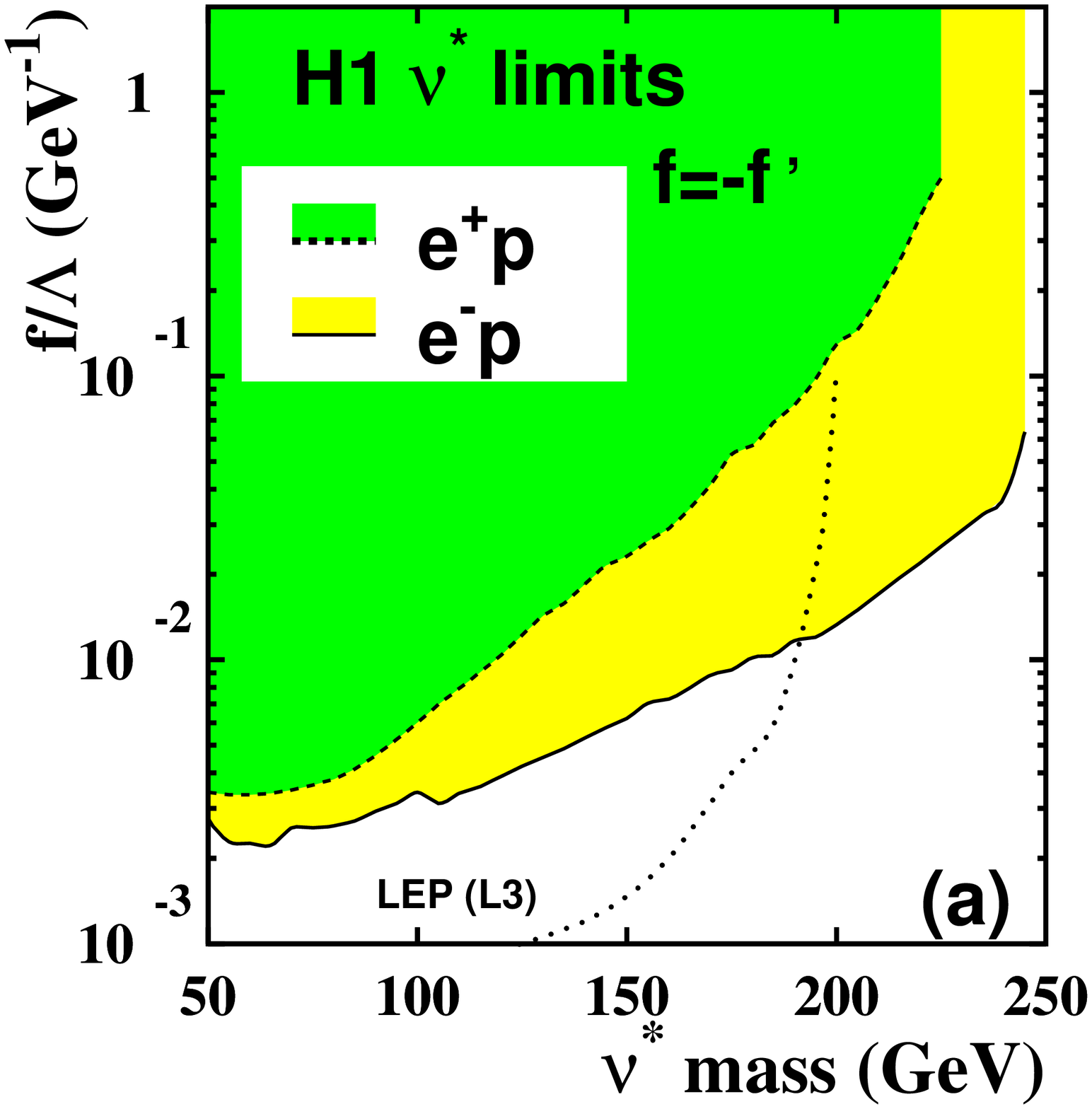,height=7.8cm}
\caption{ (left) Exclusion limits on $f/\Lambda$ as a function of $M_{e^*}$ for excited electron production. (right) Exclusion limits on $f/\Lambda$ as a function of $M_{\nu^*}$ for excited neutrino production.}\label{fig:exferm}
\end{center}
\end{figure}
The large number of quarks and leptons in the Standard Model suggests the possibility that they may be composite particles, consisting of combinations of more fundamental entities.
The observation of excited states of quarks or leptons would be a clear signal for compositeness.
The production cross-section and partial decay widths can be calculated using an effective Lagrangian~\cite{hagiwara} which depends on a compositeness mass scale $\Lambda$ and on weight factors $f$ and $f'$ describing the relative coupling strengths of the excited lepton to the $SU(2)_L$ and $U(1)_Y$ gauge bosons, respectively.

At HERA excited fermion states could be singly produced through $t$-channel gauge boson exchange. An excited electron ($e^*$) could be produced by $\gamma,Z$ exchange, whereas for $\nu^*$ production a $W$ must be exchanged. 
Note that the $\nu^*$ production cross section in $e^-p$ collisions is enhanced with respect to $e^+p$ collisions due to the quark content of the proton.
An excited fermion ($f^*$) can decay into an electron or a neutrino by radiating a gauge boson ($\gamma,W,Z$) with branching ratios determined by the $f^*$ mass ($M_{f^*}$) and the coupling parameters $f$ and $f'$.
If a fixed relationship between $f$ and $f'$ (usually $f=f'$ or $f=-f'$) is assumed, the production cross section and partial decay widths depend on two parameters only, namely the mass $M_{f^*}$  and the ratio $f/\Lambda$.

Neither collaboration has found deviations from the SM expectation in any of the considered decay channels~\cite{h1exel,h1exneu,zexel} and therefore limits have been set for $f/\Lambda$ as a function of the mass $M_{f^*}$. 
In figure~\ref{fig:exferm} (left) the HERA limits on excited electron production are shown.
For comparison the results obtained by the LEP collaborations are given as well.
The direct LEP limits are very stringent below the LEP center-of-mass energy, whereas HERA has access to higher masses.
The same is true for the limits on $f/\Lambda$ for excited neutrinos, see figure~\ref{fig:exferm} (right). 
Although the statistics for the $e^-p$ data is rather limited ($15\,\mathrm{pb}^{-1}$) this data set gives much more stringent exclusion limits than the $e^+p$ data set ($110\,\mathrm{pb}^{-1}$) due to the increased production cross section. 
Hence a substantial improvement is expected for the high luminosity runs of HERA\,II starting in autumn 2003.

\section{Isolated lepton events and FCNC single top production}
\begin{table}[t]
\begin{center}
 \begin{tabular}{|c|c|c|c|}
      \hline
      H1 $e^{\pm}p$ & Electrons & Muons & Taus\\
      \hline
      25\,GeV $<P_T^X <$ 40\,GeV & 1 / 0.95${\pm0.14}$ & 3 / 0.89$\pm0.14$ & -- \\
      $P_T^X >$ 40\,GeV & 3 / 0.54$\pm0.11$ & 3 / 0.55$\pm0.12$ & -- \\
      \hline
      \hline
      ZEUS (prelim.) $e^{\pm}p$ & Electrons & Muons & Taus\\
      \hline
      $P_T^X >$ 25\,GeV & 2 / 2.90$^{+0.59}_{-0.32}$ & 5 / 2.75$^{+0.21}_{-0.21}$ & {2} / {0.12} $\pm$ 0.02 \\
      $P_T^X >$ 40\,GeV & 0 / 0.94$^{+0.11}_{-0.10}$ & 0 / 0.95$^{+0.14}_{-0.10}$ & {1} / {0.06} $\pm$ 0.01 \\
      \hline
    \end{tabular}
\end{center}
\caption{Number of events with isolated lepton and missing transverse momentum (observed/expected).\label{tab:iso}}
\end{table}

\begin{figure}[t]
\begin{center}
\epsfig{figure=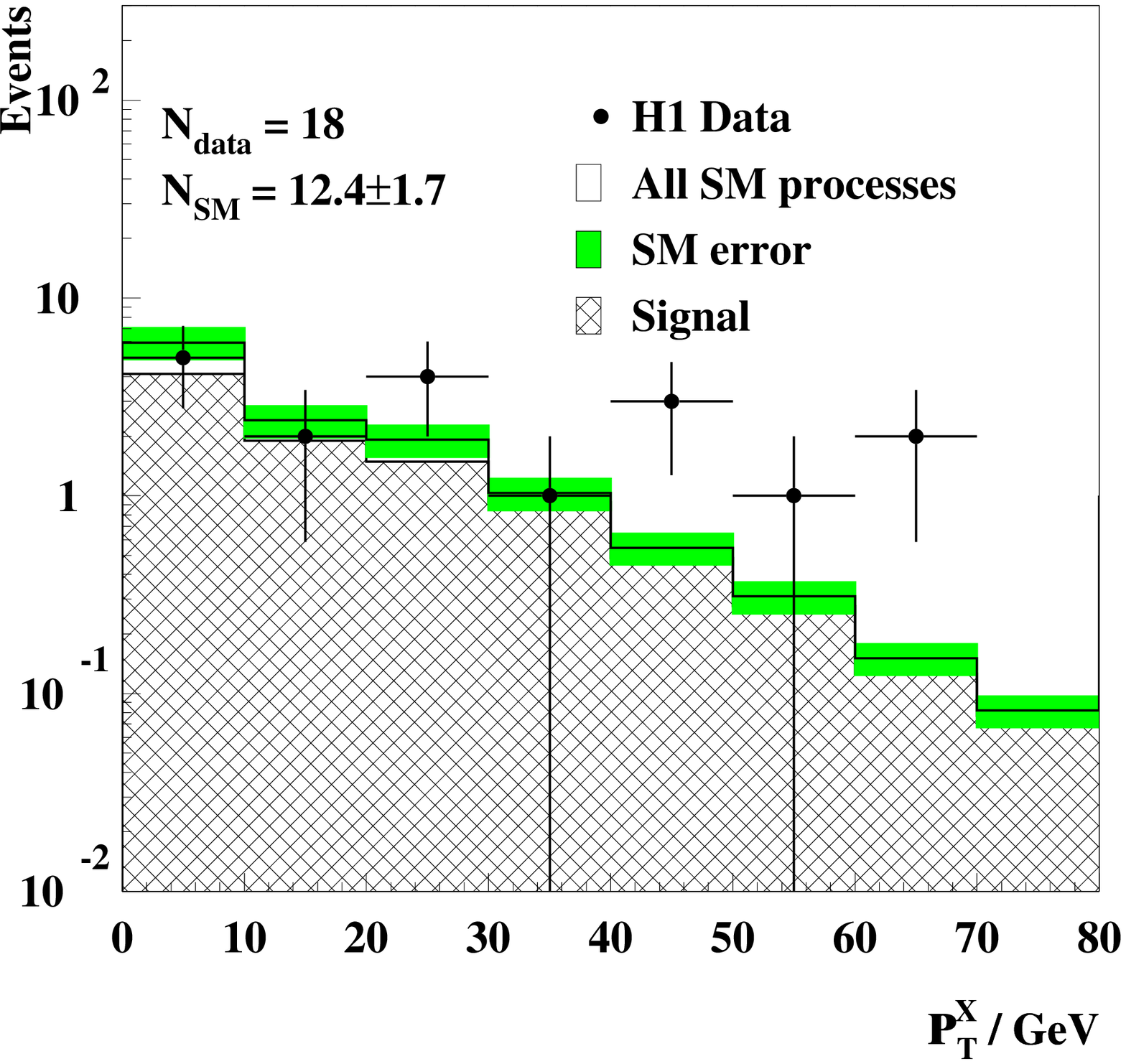,height=7.cm}
\epsfig{figure=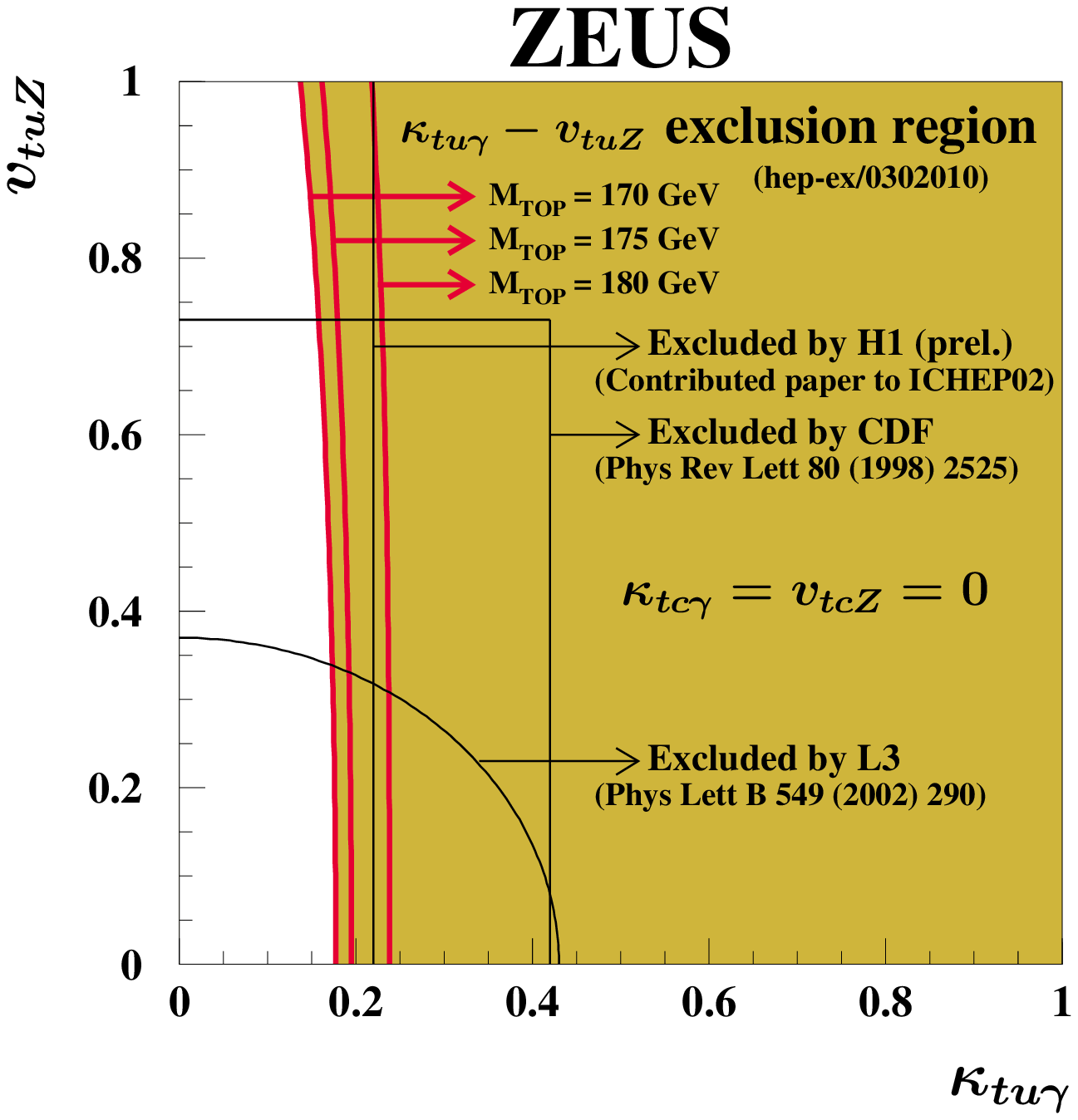,bbllx=100,bblly=170,bburx=590,bbury=590,height=7.3cm,clip=}
\caption{(left) Number of events with isolated leptons (electron and muons) and  missing transverse momentum as a function of $p_T^X$, the transverse momentum of the hadronic system. (right) Exclusion area for anomalous single top production with a FCNC vertex. Shown is the ($\kappa_{tu\gamma},v_{tuZ}$)--plane. The area on the right and above the lines is excluded.}\label{fig:top}
\end{center}
\end{figure}
In 1998 H1 reported an anomalously large number of events with a high transverse momentum lepton and missing $p_T$~\cite{98isolep}.
This analysis has now been extended to the full HERA\,I data set~\cite{h1isolep}.
The ZEUS experiment has presented a similar analysis of isolated lepton events including all events taken from 1994-2000~\cite{zeustop}.
The results, after applying two different cuts on the transverse momentum of the full hadronic state ($p_T^X$), are shown in table~\ref{tab:iso} in comparison with the SM expectation, which is dominated by $W$ production.

The number of events with isolated electrons and muons seen in H1 is in agreement with the SM for small values of $p_T^X$.
But it clearly deviates from the SM prediction for high values of $p_T^X$ and the excess is equally distributed in both channels. 
The differential distribution of these events is shown in figure~\ref{fig:top} (left).
The probability for an agreement with the SM is of the order of $10^{-3}$.
The H1 excess in the electron and muon channel is not supported by the ZEUS analysis as can be seen in table~\ref{tab:iso}. 

In addition ZEUS has performed a search for isolated taus by looking for pencil-like hadronic jets~\cite{ztau}. 
The results are given in table~\ref{tab:iso}. In this preliminary analysis two tau events are found, while only 0.12$\pm$0.02 are expected from SM processes.

Since the event topology discussed above is similiar to the topology coming from a top quark decaying semi-leptonically $t\rightarrow bW\rightarrow bl\nu$,
a possible explanation beyond the SM could be single top production $ep\rightarrow etX$.
At HERA the SM cross section for top production is tiny.
However the process could proceed via flavor changing neutral currents (FCNC) at anomalous vertices $ut\gamma$ and $tuZ$.
This FCNC transition can be parameterised by an effective Lagrangian~\cite{toptheo} which depends on the magnetic coupling $\kappa_{ut\gamma}$ and the vector coupling $v_{tuZ}$.

The top decay to hadrons has been investigated by both collaborations~\cite{zeustop,h1top}. The results are in agreement with the SM. Due to the large QCD background the sensitivity in the hadronic channel is lower with respect to the leptonic one and no stringent conclusion can be drawn in this channel.

Combining the results of the leptonic and hadronic channels limits on the anomalous couplings $\kappa_{ut\gamma}$ and $v_{tuZ}$ can be derived.
The excluded area in the ($\kappa_{ut\gamma},v_{tuZ}$)--plane is shown in figure~\ref{fig:top} (right). 
The HERA results are mainly sensitive to $\kappa_{ut\gamma}$.
The ZEUS collaboration sets more stringent limits compared to H1 due to the excess of events in the leptonic channel seen in H1.
For a final interpretation of the excess more luminosity from HERA\,II runs is needed.

\end{document}